\documentclass{elsart}
\usepackage[]{latexsym,amsmath,amssymb}
\usepackage[]{epsfig}
\newcommand{\be}{\begin{eqnarray}}
\newcommand{\ee}{\end{eqnarray}}

\begin{document}
\begin{frontmatter}
\title{\bf Thermodynamics for radiating shells in anti-de Sitter
space-time}
\author[a]{G.L.~Alberghi\thanksref{aa}}
\author[a]{R.~Casadio\thanksref{bb}}
\author[a]{G.~Venturi\thanksref{cc}}
\address[a]{Dipartimento di Fisica, Universit\`a di Bologna, and
I.N.F.N, Sezione di Bologna,
Via Irnerio 46, 40126 Bologna, Italy.}
\thanks[aa]{E-mail: alberghi@bo.infn.it}
\thanks[bb]{E-mail: casadio@bo.infn.it}
\thanks[cc]{E-mail: armitage@bo.infn.it}
\begin{abstract}
A thermodynamical description for the quasi-static collapse of
radiating, self-gravitating spherical shells of matter in
anti-de Sitter space-time is obtained.
It is shown that the specific heat at constant area and other
thermodynamical quantities may diverge before a black hole has
eventually formed.
This suggests the possibility of a phase transition occurring
along the collapse process.
The differences with respect to the asymptotically flat case
are also highlighted.
\end{abstract}
\end{frontmatter}
\section{Introduction}
\label{sec:introduction}
The collapse of spherically symmetric, self-gravitating thin shells has
been widely studied as a simplified model for the process of black hole
formation \cite{Pretorius,Giddings,Vakkuri,Trodden}.
In Ref.~\cite{Pretorius} it was shown that the black hole entropy,
expressed in terms of the area of the horizon, can be interpreted as
the entropy of a shell of matter that contracts reversibly from infinity
to its event horizon.
A thermodynamical formalism was then introduced in order to describe the
contraction of the shell.
In Refs.~\cite{Giddings,Vakkuri} the quasi-static collapse of a
non-radiating dust shell was investigated in the perspective of applying
the AdS-CFT correspondence \cite{AdSCFT} to the gravitational collapse,
as a first step with the aim of obtaining a unitary description for
the black hole formation and evaporation.
\par
In these notes, we will examine the collapse of a {\em radiating\/}
spherical shell of matter in anti-de Sitter (AdS) space-time.
The inclusion of radiation in the model may in fact help in the
understanding of the black hole formation, as suggested in
Ref.~\cite{Giddings}.
The collapse is assumed to be a {\em quasi-static\/} process, in the
sense that the shell contraction velocity is sufficiently small so
that the system (shell-matter) can be described as evolving through a
succession of equilibrium states.
This assumption  allows us to introduce a thermodynamical formalism
(see Refs.~\cite{Pretorius,Alberghi} for the case of an asymptotically
flat space-time) to describe the process.
The properties of the system depend on the equation of state, that is
a relation between the thermodynamically independent quantities.
In order to obtain some explicit results we shall consider the case of
a power-law dependence of the shell temperature (introduced as usual
through the second principle of thermodynamics) on the horizon
radius.
In Appendix~\ref{app} the particular choice corresponding to Hawking
temperature is then considered.
\par
We use units for which $\hbar=c=k_{\rm B}=1$, with $k_{\rm B}$ the
Boltzmann constant.
\section{Thermodynamics}
\setcounter{equation}{0}
\label{thermodynamics}
The spherically symmetric space-time we consider is divided into an inner
region and an outer one by a thin massive spherical shell.
The inner region can be expressed in static coordinates as
\be
ds^2_{\rm i}=-f_{\rm i}(r)\,dt^2+{dr^2\over f_{\rm i}(r)}
+r^2\,d\Omega^2
\ ,
\ee
and will be taken to be described by a Schwarzschild metric,
so that
\be
f_{\rm i}(r)=1-{2\,m\over r}
\ ,
\ee
where $m$ is a constant ADM mass.
The outer region, because of the radiation emitted by the shell,
is described by a Vaidya-AdS space-time
\be
ds^2_{\rm o}=-{1\over f_{\rm o}(r,t)}\,\left[
\left({\partial_t M(r,t)\over\partial _r M(r,t)}\right)^2
dt^2-dr^2\right]+r^2\,d\Omega^2
\ ,
\ee
with
\be
f_{\rm o}(r,t)=1-{2\,M(r,t)\over r}+{r^2\over\ell^2}
\ee
where $M(r,t)$ is the Bondi mass and its dependence on the time $t$
is related to the amount of radiation (energy) flowing out of the shell,
$\partial_t M$ and $\partial_r M$ are the partial derivatives of
$M(r,t)$ with respect to $t$ and $r$ respectively,
and $\ell$ is the AdS radius.
\par
We shall obtain the thermodynamical description for the evolution of a
thin shell, by assuming that the collapse can be described as a sequence
of equilibrium states.
Israel's junction equations \cite{Israel} for a static thin
shell located at radius $r=R$, allow us to relate the proper mass of the
shell $E$ to the inner and outer metrics through the equation
\be
E(R,M)=4\,\pi\,R^2\,\rho=
R\,\left(\sqrt{f_{\rm i}(R)}-\sqrt{f_{\rm o}(R)}\right)
\ ,
\label{energy}
\ee
where $\rho$ is the surface energy density, and to evaluate the surface
tension, denoted by $P$, as
\be
&&P(R,M)\equiv{\partial E\over \partial A}
\nonumber
\\
&&=
{1\over 8\,\pi\,R}\,\left[
\sqrt{f_{\rm i}(R)}-\sqrt{f_{\rm o}(R)}
+{1\over\sqrt{f_{\rm i}(R)}}\,{m\over R}
-{1\over\sqrt{f_{\rm o}(R)}}\,\left({M\over R}+{R^2\over \ell^2}\right)
\right]
\ ,
\label{pressure}
\ee
where $A=4\,\pi\,R^2$ is the shell area.
The continuity equation for the matter has to be taken as a constraint,
and can be expressed in the form
\be
{dL\over d\tau}={1\over\sqrt{f_{\rm o}(R)}}\,{dM\over d\tau}
\ ,
\label{continuity}
\ee
where  $L$ is the shell luminosity and $\tau$ is the proper time
of an observer sitting on the shell.
\par
In order to obtain a description of the collapse process in a
thermodynamical language one has to set up a correspondence between the
mechanical properties of the shell, such as its tension and proper mass,
and thermodynamical quantities such as the pressure, the internal energy,
the temperature or the entropy.
The proper mass of the shell is naturally identified with its internal
energy (see Refs.~\cite{Pretorius,Alberghi}) and the surface tension with
the thermodynamical pressure.
This means that the shell, considered as a thermodynamical system, is
characterized by an internal energy $E(R,M)$ and a pressure $P(R,M)$.
We note that in our formalism the Schwarzschild mass $M$ and the
radius $R$ are taken to be the dynamical independent variables,
whereas $m$ and $\ell$ are taken to be fixed parameters.
Therefore, we shall often find it convenient to use $M$ instead of the
horizon radius $R_h$, the latter being defined by $f_{\rm o}(R_h)=0$,
that is
\be
M={R_h\over 2}\,\left(1+{R_h^2\over\ell^2}\right)
\ .
\ee
\par
One may now introduce the first law of thermodynamics, associated with
energy conservation, by defining the infinitesimal heat flow $\delta Q$
by
\be
\delta Q=dE-P\,dA
\ .
\label{I}
\ee
On using the explicit expressions for the pressure and the internal energy
one finds
\be
\delta Q={dM\over\sqrt{f_{\rm o}(R)}}
\ .
\ee
This expression agrees with the luminosity of a collapsing shell as
described in Eq.~(\ref{continuity}).
The above Eq.~(\ref{I}) can thus be re-interpreted as the continuity
equation for the matter on the shell, and as such it confirms that the
definitions of the internal energy (\ref{energy}) and pressure
(\ref{pressure}) are correct.
\par
It is now possible to introduce a temperature $T$ through the second
principle of thermodynamics, that is the existence of the entropy
as the exact differential
\be
dS={\delta Q\over T}
\ .
\ee
The temperature appears as an integrating factor which must satisfy the
integrability condition
\be
{\partial\over\partial R}\,
\left(T\,\sqrt{f_{\rm o}(R)}\right)^{-1}
=0
\ ,
\ee
whose general solution is
\be
T={B_h(R_h)\over\sqrt{f_{\rm o}(R)}}
\ ,
\label{temperature}
\ee
where $B_h=B_h(R_h)$ is an arbitrary function of the horizon radius
$R_h$, leading to
\be
dS=\left(1+3\,{R_h^2\over\ell^2}\right)\,{dR_h\over 2\,B_h}
\ .
\label{dS}
\ee
We note that the temperature exhibits the usual Tolman radial dependence.
Once the temperature is fixed one may evaluate the specific heat
at constant radius $C_R$,
\be
C_R&\equiv&
T\,\left({\partial S\over \partial T}\right)_R
=
T\,\left({\partial S\over \partial R_h}\right)_R\,
\left({\partial T\over\partial R_h}\right)_R^{-1}
\nonumber
\\
&=&
\left[{2\,\ell^2\,B_h'\over \ell^2+3\,R_h^2}
+{B_h\over R\,f_{\rm o}(R)}\right]^{-1}
\ ,
\ee
where $B_h'=dB_h/dR_h$.
The above expression shows a possible singularity for $R$
satisfying
\be
\left(R-R_h\right)\,\left(\ell^2+R^2+R\,R_h+R_h^2\right)
=-{\ell^2+3\,R_h^2\over\left(\ln B_h^2\right)'}
\ .
\label{singularity1}
\ee
\par
The specific heat at constant tension takes the form
\be
C_P&=&T\left({\partial S\over \partial T}\right)_P
\nonumber
\\
&=&
{T\over 2\,B_h}\left(1+3\,{R_h^2\over\ell^2}\right)\,
\left[\left({\partial T\over \partial R_h}\right)_R
-\left({\partial T\over \partial P}\right)_{R_h}\,
\left({\partial P\over \partial R_h}\right)_R\right]^{-1}
\ ,
\ee
whose explicit expression we omit for the sake of brevity.
Other thermodynamical quantities of interest, related to the second
derivative of the Gibbs potential \cite{Zemanski}, are the change
in area with respect to the temperature for fixed tension
$\left({\partial A/\partial T}\right)_P$ and with respect to the
tension for fixed temperature $\left({\partial A/\partial P}\right)_T$.
All such quantities show a singular behavior if there exists an $R$
satisfying
\be
&&
{3\,R_h\over 2\,R}\,\left(1+{R_h^2\over\ell^2}\right)\,
\left[1-{R_h\over 2\,R}\,\left(1+{R_h^2\over\ell^2}\right)
-{R^2\over\ell^2}\right]
\nonumber
\\
&&+\left[{f_{\rm o}(R)\over f_{\rm i}(R)}\right]^{3/2}
\left(1-{3\,m\over R}+{3\,m^2\over R^2}\right)
\nonumber
\\
&&=1-{R_h^2\over 4\,R^2}\,
\left[1+{R_h^2\over\ell^2}+{2\,R^3 \over R_h\,\ell^2}\right]^2
\left[1+f_{\rm o}(R)\,
{\ell^2\,\left(\ln B_h^2\right)'\over \ell^2+3\,R_h^2}\right]^{-1}
\ .
\label{singularity2}
\ee
\par
\begin{figure}
\centerline{
\epsfxsize=160pt\epsfbox{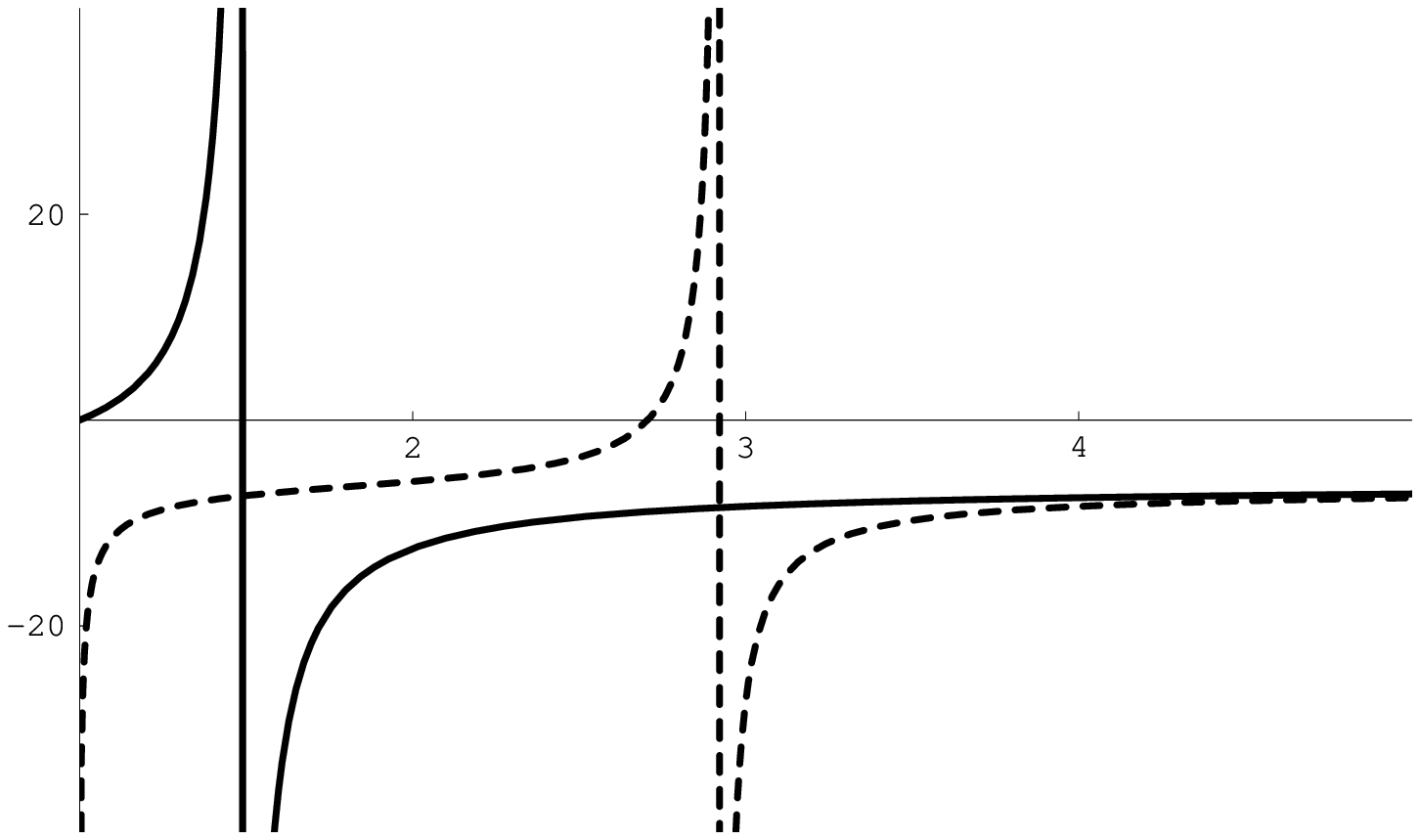}
\raisebox{1.6cm}{${r\over R_h}$}
\hspace{0.5cm}
\epsfxsize=160pt\epsfbox{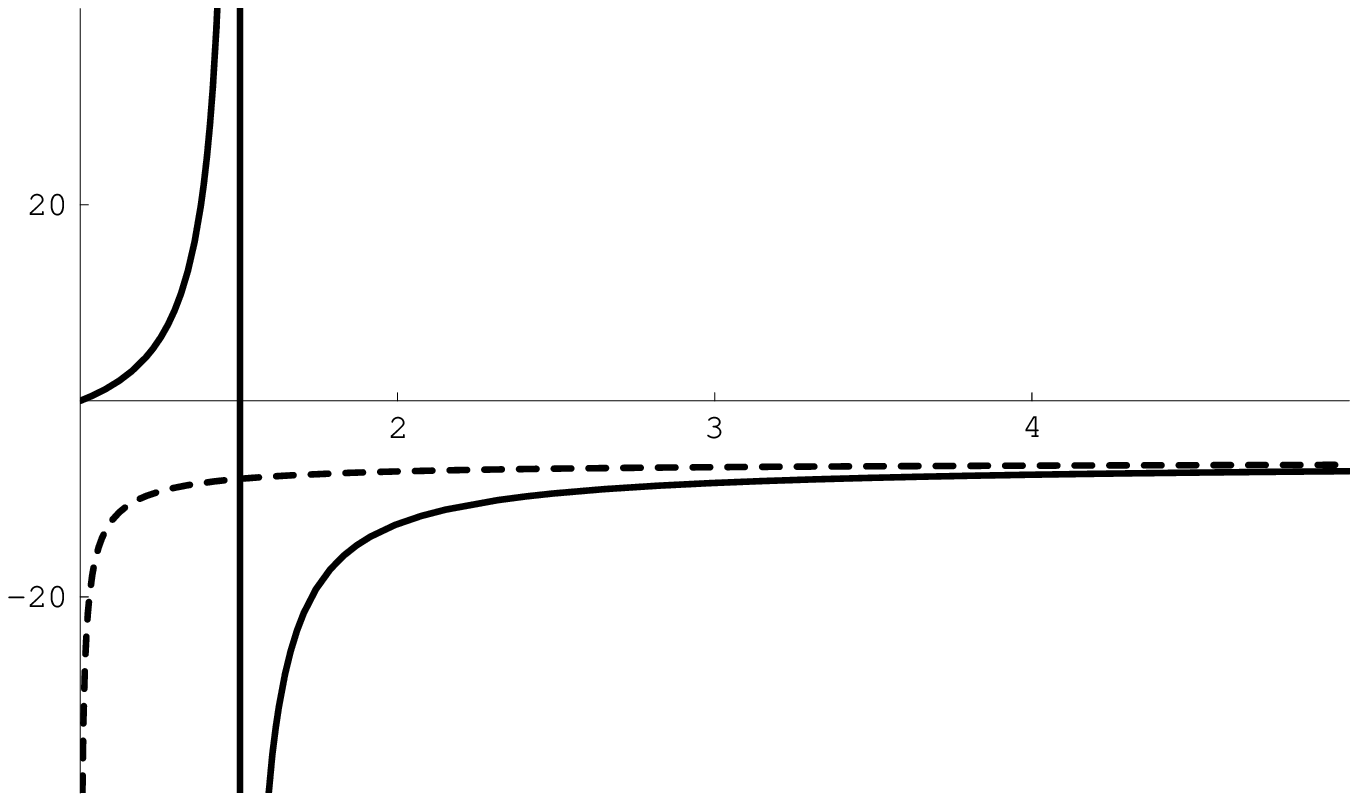}
\raisebox{1.6cm}{${r\over R_h}$}
}
\par
\noindent
\hspace{4cm}{\bf (a)}\hspace{7cm}{\bf (b)}
\caption{Behaviour of the specific heats at constant
area $C_R$ (solid line) and constant tension $C_P$ (dashed line)
for $m=0$, $a=1$ and $R_h=1$:
{\bf (a)} in AdS with $\ell=10$; and {\bf (b)} in asymptotically
flat space ($\ell\to\infty$).}
\label{grafico1}
\end{figure}

In order to have an explicit expression for the specific heats and to
proceed further in our investigation, we need an equation of state, that
is an expression for the function $B_h$.
Let us examine a rather general case assuming a power-law dependence of
the function $B_h$ on the horizon radius, leading to the temperature
\be
T={1\over \sqrt{f_{\rm o}(R)}}\,{1\over 4\,\pi\,R_h^a}
\ ,
\label{pow}
\ee
with $a$ a constant.
We can now determine the specific heat at constant area
\be
C_R=
-4\,\pi\,f_{\rm o}(R)\,R_h^{a+1}\,
\left(1+3\,{R_h^2\over\ell^2}\right)\,
\left(2\,a\,f_{\rm o}(R)+R_h\,{\partial f_{\rm o}(R)\over\partial R_h}
\right)^{-1}
\ .
\ee
This implies that $C_R $ diverges for
\be
0&=&
2\,a\,f_{\rm o}(R)+R_h\,{\partial f_{\rm o}(R)\over\partial R_h}
\nonumber
\\
&=&
2\,a\,\left(1+{R^2\over \ell^2}\right)
-{R_h\over R}\,\left[(1+2\,a)+(3+2\,a)\,{R_h^2\over \ell^2}\right]
\ .
\label{CR0}
\ee
Let us examine the above equation.
One finds that $C_R$ has at most one singularity at $R=R_\ell$ for
\be
\begin{array}{lcc}
\left\{
\begin{array}{l}
a>0
\\
\\
\forall\,R_h\ge 0
\end{array}
\right.
& {\rm with} & R_\ell>R_h
\ ,
\end{array}
\label{Rl>}
\ee
and
\be
\begin{array}{lcc}
\left\{
\begin{array}{l}
-3/2\le a \le -1/2
\\
\\
R_h^2\le-\displaystyle{1+2\,a\over 3+2\,a}\,\ell^2
\end{array}
\right.
&
{\rm with} & 0\le R_\ell<R_h
\ .
\end{array}
\label{Rl<}
\ee
The behaviour of the specific heat at constant tension $C_P$ is
singular for $a>-3/2$, as is shown (along with $C_R$) in
Fig.~\ref{grafico1}, for any value of $\ell$, and the radius for
which the singularity appears decreases as $a$ increases.
\par
It is now interesting to compare the above singular behaviours with
that of the asymptotically flat case, which is obtained for
$\ell\to\infty$.
For instance, Eq.~(\ref{CR0}) simplifies considerably and one finds
that the singularity moves to
\be
R_{\infty}=\left(1+{1\over 2\,a}\right)\,R_h
\ ,
\label{CRinf}
\ee
which is a physical (i.e.~positive) radius for $a>-1/2$
and is larger than $R_h$ for $a>0$.
Let us note that the singularity (\ref{Rl<}), which occurs inside the
shell horizon in AdS, is now replaced by an again hidden (inside the
horizon) one for $-1/2<a<0$.
\section{Conclusions}
\setcounter{equation}{0}
\label{conclusions}
We have analysed the thermodynamical behaviour for the collapse of a
radiating shell in an AdS space-time, under the assumption that the
evolution consists of a succession of equilibrium states, that is the
process is quasi-static.
On identifying the internal energy and surface tension of the shell,
we were able to evaluate the specific heats at constant area and tension
and other related thermodynamical quantities when the temperature is
given by a power law of the horizon radius as in Eq.~(\ref{pow}).
Their behavior may suggest the existence of a phase transition before
the shell reaches its Schwarzschild radius.
Of course, in a realistic case, the shell ADM mass and horizon could
change quickly in time and the adiabaticity (quasi-staticity) of the
process may be lost.
There is however evidence for cases in which the shell naturally
emits radiation (having a Hawking temperature) in such a way that
its contraction velocity remains small \cite{acvv} and the quasi-static
approximation can therefore be applied.
\par
The case of the Hawking temperature is not of the form (\ref{pow})
and is analyzed in Appendix~\ref{app}.
A very interesting feature of the model is then the appearance of a
threshold value for the AdS parameter $\ell\simeq 7\,R_h/4$, which
leads to two very different behaviours for the specific heats
at constant area and tension, as shown in Fig.~\ref{grafico2}.
\par
We feel that these singularities in thermodynamically quantities such
as specific heats may be of relevance and deserve further investigation.
\appendix
\section{Hawking temperature}
\setcounter{equation}{0}
\label{app}
Let us examine the case in which the temperature is that of a black hole
with horizon radius $R_h$ \cite{hawking}, implying
\be
B_h={1\over 4\,\pi\,R_h}\,\left(1+3\,{R_h^{2}\over \ell^{2}}\right)
\ ,
\label{hawkingT}
\ee
which seems to be the most natural choice if we assume that at the
end of the collapse the system behaves as if a black hole were being
formed (for an analysis supporting the naturalness of this choice see
Refs.~\cite{Alberghi,Oppenheim,Belgiorno}).
On substituting for $B_h$ in Eq.~(\ref{singularity1}) one
obtains the equation
\be
R+{R^{3}\over \ell^{2}}=
{R_h\over 2}\,\left(3+2\,{R_h^{2}\over\ell^{2}}+3\,{R_h^{4}\over\ell^{4}}
\right)
\ ,
\ee
which determines the singularity of the specific heat at constant area.
Let us note that for $\ell\to\infty$, the singularity for
the specific heat at constant radius is located at $R=3\,M$
as in the asymptotically flat case [see Eq.~(\ref{CRinf}) and
Ref.~\cite{Alberghi}].
\par
In order to examine the singularities of $C_R$ and $C_P$
for a general value of $\ell$ one must study Eqs.~(\ref{singularity1})
and (\ref{singularity2}).
This analysis shows that for $\ell>\ell_0\simeq 7\,R_h/4$,
$C_R$ has a singularity and changes sign for a finite radius, as shown
in Fig.~{\ref{grafico2}.
As $\ell$ approaches $\ell_0$ from above, the singularity
moves to arbitrarily large values of $R$.
On the other hand the specific heat at constant tension $C_P$,
shows a singularity at the horizon $R_h$ and at a finite radius.
The singularity moves to arbitrarily large radii as $\ell\to\ell_0$
and $\ell\to\infty$.
The singularity of $C_P$ always occurs for a radius greater than that
for which $C_R$ is singular.
\par
For the case $\ell<\ell_0$, $C_R$ does not show any singularity and
remains positive for any radius of the shell, becoming zero at the
horizon radius, whereas $C_P$ is regular everywhere except at the horizon,
as is shown in Fig.~\ref{grafico2}.
It is worth noting that the singularity in $C_P$ is not present in the
purely Schwarzschild case, thus it is a peculiar feature for the AdS
space-time.
\par
We finally note that for the choice of a Hawking temperature,
the entropy, following Eqs.~(\ref{dS}) and (\ref{hawkingT}),
is given by
\be
S=\int {\delta Q\over T}=\pi\,R_h^2
={1\over 4}\,({\rm horizon\ area})
\ .
\label{SA}
\ee
This expression will exhibit a simple additive property, in the
sense that the entropy of two non-interacting (well separated)
shells will just be the sum of the two entropies, as expected
for usual thermodynamical systems \cite{Belgiorno}.
Such an additive property also rules out any integration constant
in the Eq.~(\ref{SA}).
\begin{figure}
\centerline{
\epsfxsize=160pt\epsfbox{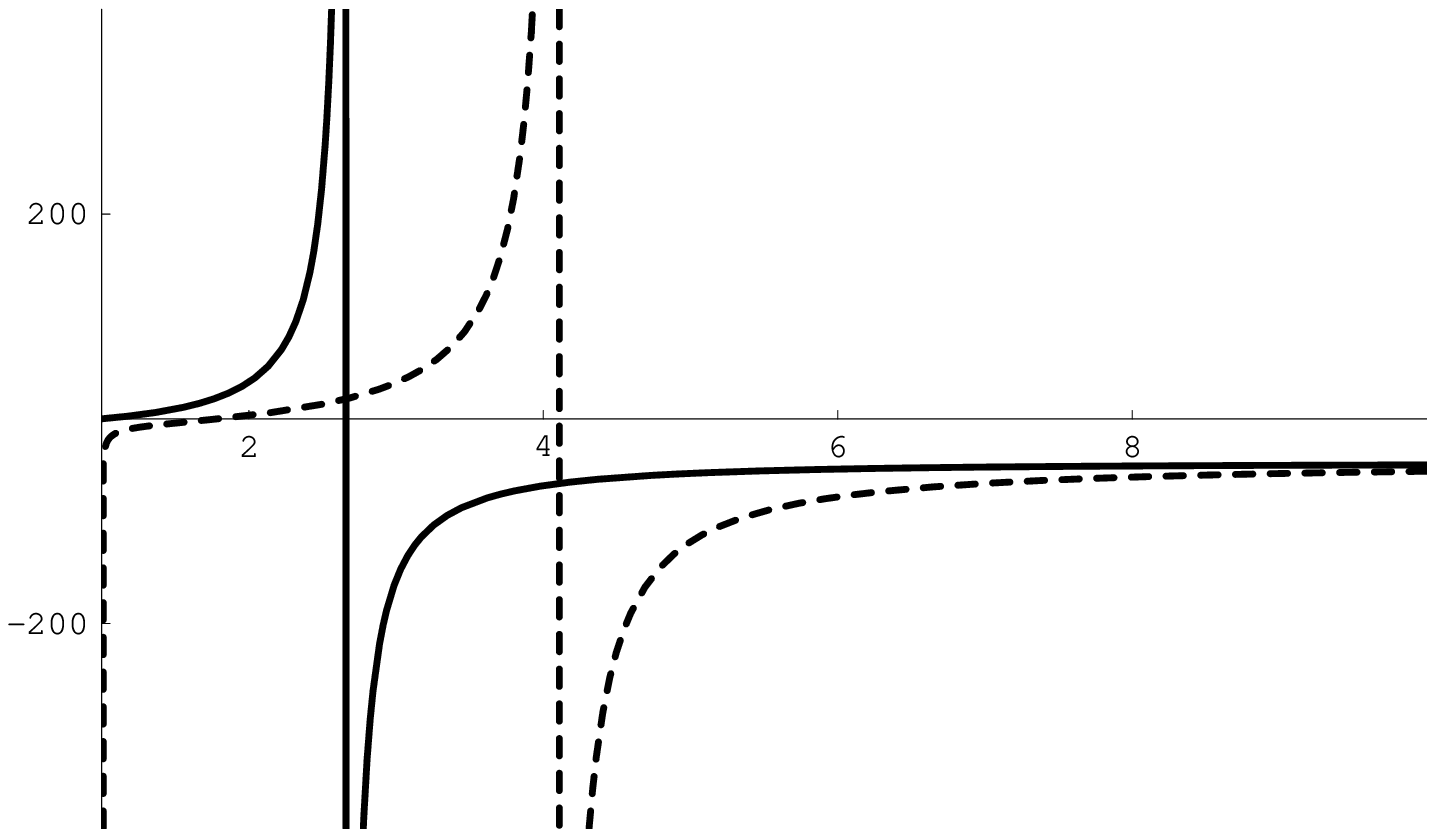}
\raisebox{1.6cm}{${r\over R_h}$}
\hspace{0.5cm}
\epsfxsize=160pt\epsfbox{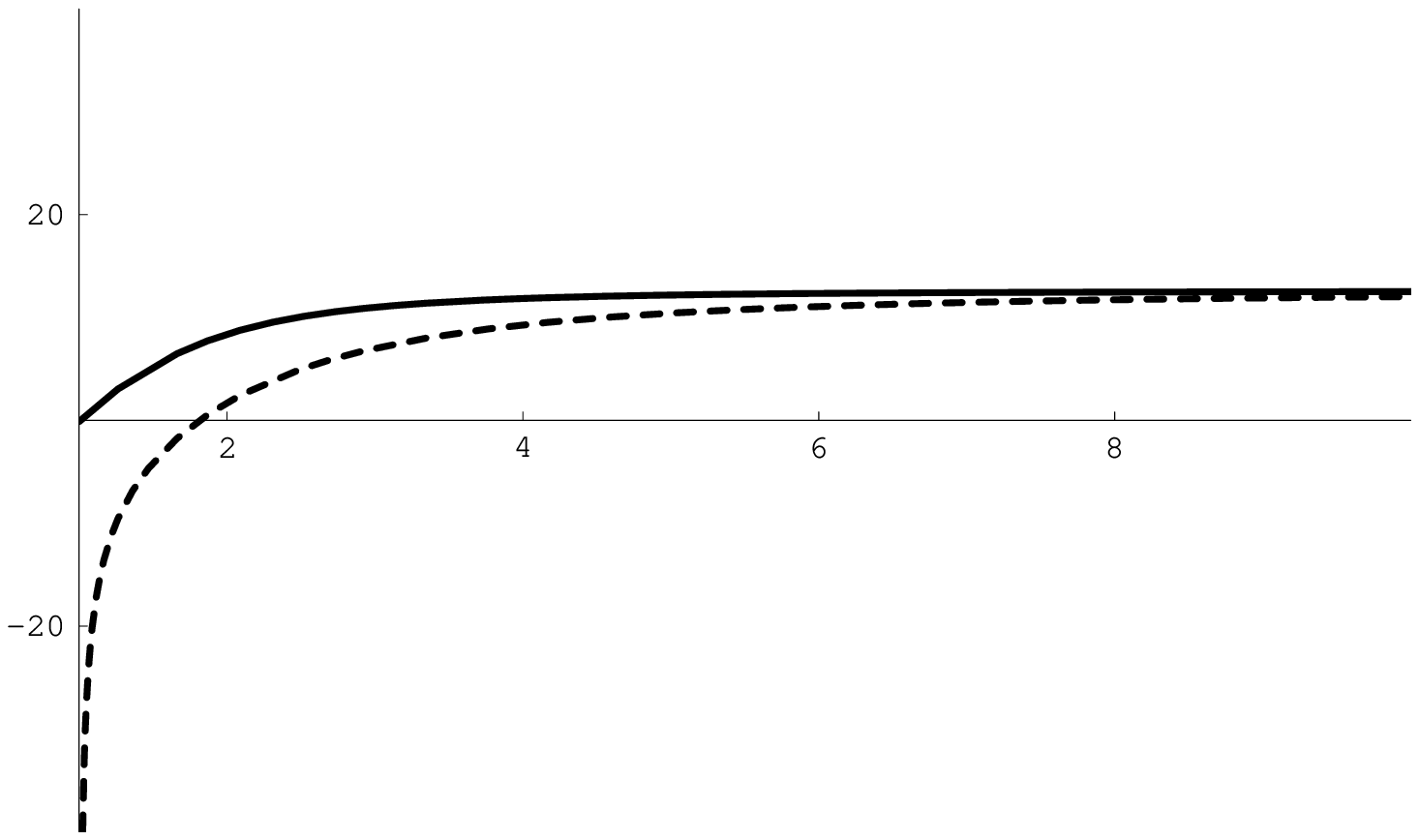}
\raisebox{1.6cm}{${r\over R_h}$}
}
\par
\noindent
\hspace{4cm}{\bf (a)}\hspace{7cm}{\bf (b)}
\caption{Behaviour of the specific heats at constant
area $C_R$ (solid line) and constant tension $C_P$ (dashed line)
for $m=0$ and $R_h=1$: {\bf (a)} $\ell=2\,R_h>\ell_0$; and
{\bf (b)} $\ell=R_h<\ell_0$.}
\label{grafico2}
\end{figure}
\end{document}